\documentclass[a4paper]{memsissa} 
\usepackage{natbib} 
\usepackage{graphicx} 

\idline{0}{0} 
\begin{document} 
\title{BASTI: An interactive database of updated stellar evolution models s 
} 
 
\author{ S. Cassisi\inst{1}, A. Pietrinferni\inst{1,2}, 
M. Salaris\inst{3}, F. Castelli\inst{4}, D. Cordier\inst{5}, 
\and M. Castellani\inst{6}\fnmsep 
}

\offprints{S. Cassisi} 
\mail{INAF - Astronomical Observatory of Collurania, 
via M. Maggini, 64100 Teramo} 
 
\institute{INAF - Astronomical Observatory of Collurania, 
via M. Maggini, 64100 Teramo, Italy \email{adriano,cassisi@te.astro.it}\\ 
\and 
Universit\'a di Teramo, viale F. Crucioli 64100 Teramo, Italy\\ 
\and  
Astrophysics Research Institute, Liverpool John Moores University, Twelve 
Quays House,Birkenhead, CH41 
1LD, UK, \email{ms@astro.livjm.ac.uk}\\ 
\and 
INAF - Astronomical Observatory of Trieste, 
via Tiepolo 11, 34131 Trieste, Italy, \email{castelli@ts.astro.it}\\ 
\and 
Ecole Nationale Superieure de Chimie de Rennes, Campus de Beaulieu, 
35700 Rennes, France, \email{daniel.cordier@ensrc-rennes.fr}\\ 
\and 
INAF - Astronomical Observatory of Rome, Via di Frascati, 33 
00040 Monte Porzio Catone, Italy, \email{m.castellani@mporzio.astro.it}\\}

\abstract{We present a new database of stellar evolution models for a 
large range of masses and chemical compositions, based on an up-to-date theoretical framework. 
We briefly discuss the physical inputs and the assumptions 
adopted in computing the stellar models. 
We explain how to access to the on-line archive and briefly discuss the interactive
WEB tools that can be used to compute user-specified 
evolutionary tracks/isochrones/luminosity functions.
The future developments of this database are also outlined.
 
} 
\authorrunning{S. Cassisi et al.} 
\titlerunning{Stellar evolution models} 
\maketitle 
%
 
\section{Introduction} 
 
A large database of stellar evolution models, spanning a wide range of stellar 
masses and initial chemical compositions, represents a fundamental tool to 
investigate the properties of stellar populations in both 
Galactic and extragalactic systems. 
 
A good - reliable - library of stellar models has to fulfill some  
important requirements: 
 
\begin{itemize} 
\item{accuracy - the physical inputs have to be updated;} 
\item{homogeneity - all the models have to be computed using the same evolutionary code and 
the same physical framework;} 
\item{completeness - for each fixed chemical composition, the range of star masses has to  
be sampled with a mass spacing appropriate to adequately cover all evolutionary stages;} 
\item{reliability - the models have to reproduce as many empirical constraints as possible;} 
\item{easy access - all results have to be easily available to the
potential users.} 
\end{itemize} 
 
We have accounted for all of these criteria to set up the archive of stellar 
evolutionary models described in this paper. 
In the next section we shortly discuss the physical inputs adopted 
in the model computation; the grid of masses and chemical compositions is discussed 
in section~3. A presentation of the WEB database interface will 
close the paper. We refer the reader interested in assess the level of
agreement between our library and observational constraints to the following  papers: 
Pietrinferni et al. (2004), Salaris et al. (2004), Cassisi et al. (2003) and references 
therein. 
 
\section{The theoretical framework} 
 
Our model database has been computed by using 
a recent version of the FRANEC evolutionary code, updated in many 
aspects concerning both the numerical scheme for treating the 
nuclear burnings and the accuracy of the numerics. Almost all the 
adopted physical inputs have been updated as well. In particular, 
the radiative opacity tables (Iglesias \& Rogers~\cite{iglesias} and  
Alexander \& Ferguson \cite{alexander}), thermal conduction (Potekhin  
\cite{pot}), plasma-neutrino processes (Haft et al.~\cite{haft}).  
The nuclear reaction rates have been updated by using the NACRE 
compilation (Angulo et al.~1999), with the exception of the 
$^{12}$C$(\alpha,\gamma)^{16}$O reaction (Kunz et al.~\cite{kunz}). 
As for the Equation of State (EOS), we employ the new EOS  
by A. Irwin\footnote{More informations about this new EOS can be found at the
following URL site http://freeeos.sourceforge.net.} (see also Cassisi,  
Salaris \& Irwin 2003), which covers all relevant evolutionary phases 
from the Main Sequence to the initial phases of White Dwarf cooling or 
advanced burning stages, for a large mass range.  
All models have been computed by fixing the extension of the
convective core during the core H-burning phase 
both classically (Schwarzschild criterion) and considering a
non-negligible efficiency of the overshoot process ($\lambda_{OV}=0.2 Hp$). 
We have also accounted for mass loss by using the Reimers (1975) formula with the 
free parameter $\eta$ set to 0.2 and 0.4. 
A more detailed discussion about the physical inputs can be found in 
Pietrinferni et al.~(2004). 
 
\section{The archive content} 
 
With the aim of covering a wide range of chemical compositions, we provide  
models for 11 different metallicities, namely 
$Z=0.0001$, 0.0003, 0.0006, 0.001, 0.002, 0.004, 0.008, 0.01, 0.0198,  
0.03 and 0.04, assuming two different heavy element distributions: scaled-solar (Grevesse \& 
Noels~1993) and $\alpha-$enhanced (Salaris \& Weiss~1998). As for the initial He-abundance, 
we adopt the value ($Y=0.245$) provided by Cassisi et
al.~\cite{csi03}; to reproduce the initial solar He-abundance obtained
from the calibration of the solar model we assume an  
Helium enrichment law equal to $\Delta{Y}/\Delta{Z}\approx1.4$. 
 
For each fixed chemical composition we have 
computed models in the mass 
range $0.5\le{M/M_\odot}\le10$ with a very fine mass spacing (see details in Pietrinferni, 
et al.\cite{pietr}). 
All models, with the exception of the least massive ones whose central H-burning time scale 
is longer than the Hubble time, have been evolved from the Pre-Main Sequence phase up to  
the C-ignition, or until the first thermal pulses along the asymptotic giant branch. 
The main characteristics of our archive are listed in table~1.  
These models have been used to compute isochrones\footnote{The interested 
user can directly download from our WEB site both evolutionary tracks and isochrones 
as single files, or as a tar gzipped archive file} for a 
wide range  of ages, from 30~Myr to the upper limit listed in Table 1. 

\begin{table*} 
\centering 
\begin{tabular}{|c|c|c|c|c|c|c|c|c|}

\hline 
\multicolumn{1}{|c|}{ {mixture}} & 
\multicolumn{4}{|c|}{ {scaled-solar}} & \multicolumn{4}{|c|}{ 
{$\alpha$-enhanced}} \\ 
 
\hline 
\multicolumn{1}{|c|}{ {$\eta$}} & 
\multicolumn{2}{|c|}{ {0.2}} & \multicolumn{2}{|c|}{ {0.4}} &  
\multicolumn{2}{|c|}{ {0.2}} & \multicolumn{2}{|c|}{ {0.4}}  \\ 
 
\hline 
\multicolumn{1}{|c|}{{$\lambda_{OV}$}} & 
\multicolumn{1}{|c|}{ {0}} & \multicolumn{1}{|c|}{ {0.2}} &  
\multicolumn{1}{|c|}{ {0}} & \multicolumn{1}{|c|}{ {0.2}} &  
\multicolumn{1}{|c|}{ {0}} & \multicolumn{1}{|c|}{ {0.2}} &  
\multicolumn{1}{|c|}{ {0}} & \multicolumn{1}{|c|}{ {0.2}} \\ 
 
\hline 
 
\multicolumn{1}{|c|}{{$N^O$} tracks} & 
\multicolumn{1}{|c|}{ {20}} & \multicolumn{1}{|c|}{ {20}} &  
\multicolumn{1}{|c|}{ {40}} & \multicolumn{1}{|c|}{ {20}} &  
\multicolumn{1}{|c|}{ {20}} & \multicolumn{1}{|c|}{ {20}} &  
\multicolumn{1}{|c|}{ {40}} & \multicolumn{1}{|c|}{ {20}} \\ 
 
\hline 
 
\multicolumn{1}{|c|}{{$M_{min}$}(M$_\odot$)} & 
\multicolumn{1}{|c|}{ {0.5}} & \multicolumn{1}{|c|}{ {1.1}} &  
\multicolumn{1}{|c|}{ {0.5}} & \multicolumn{1}{|c|}{ {1.1}} &  
\multicolumn{1}{|c|}{ {0.5}} & \multicolumn{1}{|c|}{ {1.1}} &  
\multicolumn{1}{|c|}{ {0.5}} & \multicolumn{1}{|c|}{ {1.1}} \\ 
 
\hline 
 
\multicolumn{1}{|c|}{{$M_{max}$}(M$_\odot$)} & 
\multicolumn{1}{|c|}{ {2.4}} & \multicolumn{1}{|c|}{ {2.4}} &  
\multicolumn{1}{|c|}{ {10}} & \multicolumn{1}{|c|}{ {10}} &  
\multicolumn{1}{|c|}{ {2.4}} & \multicolumn{1}{|c|}{ {2.4}} &  
\multicolumn{1}{|c|}{ {10}} & \multicolumn{1}{|c|}{ {10}} \\ 
 
\hline 
 
\multicolumn{1}{|c|}{{$N^O$} isoc.} & 
\multicolumn{1}{|c|}{ {63}} & \multicolumn{1}{|c|}{ {44}} &  
\multicolumn{1}{|c|}{ {54}} & \multicolumn{1}{|c|}{ {44}} &  
\multicolumn{1}{|c|}{ {63}} & \multicolumn{1}{|c|}{ {44}} &  
\multicolumn{1}{|c|}{ {54}} & \multicolumn{1}{|c|}{ {44}} \\ 
 
\hline 
 
\multicolumn{1}{|c|}{{$Age_{min}$}(Myr)} & 
\multicolumn{1}{|c|}{ {30}} & \multicolumn{1}{|c|}{ {30}} &  
\multicolumn{1}{|c|}{ {30}} & \multicolumn{1}{|c|}{ {30}} &  
\multicolumn{1}{|c|}{ {30}} & \multicolumn{1}{|c|}{ {30}} &  
\multicolumn{1}{|c|}{ {30}} & \multicolumn{1}{|c|}{ {30}} \\

\hline 
 
\multicolumn{1}{|c|}{{$Age_{max}$}(Gyr)} & 
\multicolumn{1}{|c|}{ {19}} & \multicolumn{1}{|c|}{ {9.5}} &  
\multicolumn{1}{|c|}{ {14.5}} & \multicolumn{1}{|c|}{ {9.5}} &  
\multicolumn{1}{|c|}{ {19}} & \multicolumn{1}{|c|}{ {9.5}} &  
\multicolumn{1}{|c|}{ {14.5}} & \multicolumn{1}{|c|}{ {9.5}} \\ 
 
\hline 
\multicolumn{1}{|c|}{{Color-$T_{eff}$}} & 
\multicolumn{8}{|c|}{{UBVRIJKL - ACS HST}} \\ 
\hline 
 
\end{tabular} 
\caption{The main characteristics of the BASTI evolutionary models database}   
\label{tab1} 
\end{table*} 

For each chemical composition we also have computed additional He-burning models  
with He core mass and envelope chemical profile fixed by a Red Giant 
Branch (RGB) progenitor having an age of $\sim13$~Gyr at the RGB tip,
and  a range of values of the total stellar mass. 
These Horizontal Branch (HB) models ($\sim30$ for each chemical
composition) constitute a valuable tool to perform synthetic HB
modeling, and to investigate pulsational and evolutionary properties
of different kinds of pulsating variable stars. 
 
All evolutionary results have been transferred from the theoretical
plane to magnitudes and colours in various photometric filters, 
by using colour-$T_{eff}$ transformations and bolometric corrections 
based on an updated set of model atmospheres (see Castelli \& Kurucz~2003 for more
details). 
At the moment, the evolutionary results are available in the
photometric filters listed in 
table~1; in the near future we will also provide models in the Str{\"o}emgren, Walraven and HST 
WFI photometric filters. It is worth to point out that, for the first time, we have these 
transformations for both scaled-solar and $\alpha-$enhanced
mixture\footnote{At present, the 
transformations for the ACS HST filters are available only for a scaled-solar chemical composition.}. 
 
We provide also tables (both in ascii and html format) for each chemical composition with  
relevant data about the theoretical models and the loci in the H-R diagram corresponding to 
the Zero Age Horizontal Branch and to the central He exhaustion.

\begin{figure*}[t]
\hspace{2cm}
\rotatebox{0}{\resizebox{11cm}{9.0cm}{
\includegraphics{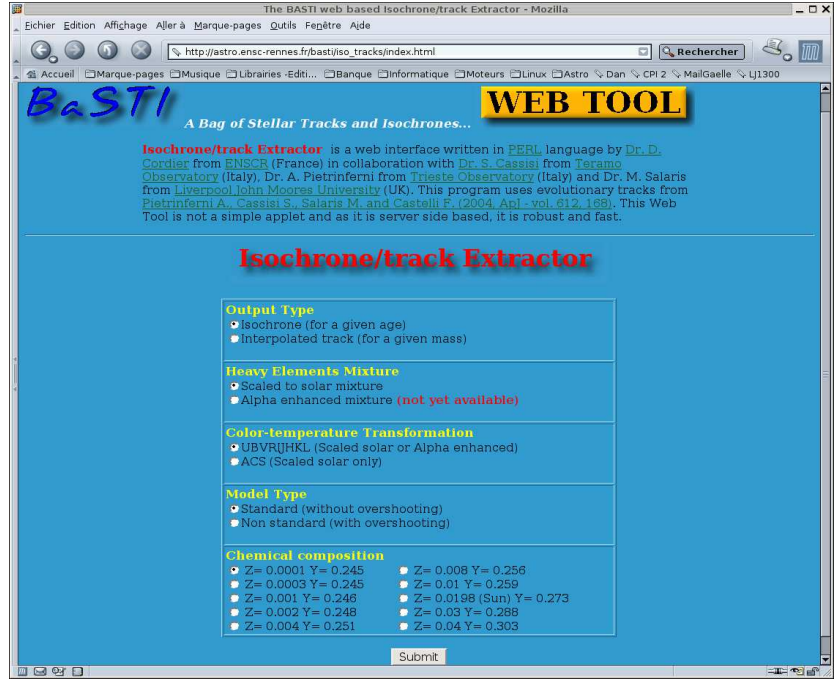}}}
\caption{\label{webtool1}The \textit{isochrones-tracks maker} web interface.}
\end{figure*} 

\section{Web library interface} 
 
The simplest way to obtain the data stored in the database is by 
using the World Wide Web. 
By following the link www.te.astro.it/BASTI/index.php the user can access our on-line 
library that includes two main sessions: the data archive and the web tools.
Needless to say, both sessions are based on the same evolutionary tracks.
 
The data archive contains all our computations listed by chemical composition. 
In this session, tracks, isochrones, HB models and various tables can be visualized  
and also downloaded.  

The web tools session is a set of three web interfaces written in
P.E.R.L.\footnote{Practical Extraction and Report Language}. Behind these interfaces there are
FORTRAN programs already used in previous published works by our
group. 
This method is powerful and
reliable because it is server side based and it uses well known FORTRAN programs.
These web tools (see figure~1) are:

\begin{itemize}
\item{isochrones-tracks maker. Using this tool it is possible to calculate 
an isochrone/track for a given age/mass for each chemical composition. The user can
also choose: heavy element mixture, color-temperature transformation and the model
type (with or without overshooting). This tool does not require registration and results
are directly sent to the Internet browser.}

\item{luminosity function maker. This tool provides the luminosity functions for a
set of isochrones previously downloaded by the user.
It is possible to select: heavy element mixture, the photometric filter, the number 
of isochrones, the number of the stars in the simulation, the and the Initial Mass
Function exponent value.
It is important to stress the fact that \textbf{this} program runs correctly only using
isochrones of this database. As in the case of the isochrones-tracks
maker, results are sent to the user's browser.}

\item{synthetic color-magnitude diagrams maker. This tool will be
available soon\footnote {A pre-release version
can be seen at http://astro.ensc-rennes.fr/basti/synth\_pop/\ index.html}. It
will allow stellar synthetic populations calculations, including
accurate estimates of 
the pulsational
properties of the expected population(s) of variable stars. 
The user will be free to fix various
parameters like photometric and spectroscopic errors, colour excess, 
fraction of unresolved binaries, etc.
This third tool will require registration
\footnote{Pre-registration can be done by sending an e-mail to 
S. Cassisi or D. Cordier}, the results being sent to the user by e-mail.}

\end{itemize}

\section{Final remarks}

This database is continuously updated by including additional stellar
evolution data. More in detail, we will soon make available
all the models (both tracks and
isochrones) in additional photometric systems like the
Str{\"o}emgren, Walraven and the HST WFI
system. We plan also to extend as soon as possible the covered
stellar mass range, adding models for both very low mass objects, and 
more massive stars than the current upper limits. In addition, we will
extend the  asymptotic giant branch evolution to cover 
the full thermal pulses' phase.

We wish to mention that we are available to compute (within one
or two weeks) - on request - any specified 
evolutionary result which is not yet included in the archive.

In the future we will also allow users to run our evolutionary code
through a simple World Wide Web browser, in order to compute ad-hoc stellar
models for a specific project.
 
\begin{acknowledgements} 

We wish to warmly thank V. Castellani for strongly pointing out the importance 
of \lq{building}\rq\ this kind of
data archive as well as for his constant encouragement. We wish also
to acknowledge in advance all people who will be using 
the BASTI archive for their own research and who will send us their comments/suggestions.
\end{acknowledgements} 
 
\bibliographystyle{aa}

\end{document}